\documentclass[aps,prl,superscriptaddress,twocolumn,groupedaddress,showpacs,amsmath,amsfonts]{revtex4}
\usepackage{graphicx}
\usepackage{bm}

\begin {document}
\title {Achievement of a record electron temperature for a magnetic mirror device}

\author{P.\,A.\,Bagryansky}
\affiliation{Budker Institute of Nuclear Physics, 11 Lavrentieva prospect, 630090 Novosibirsk, Russia} 
\affiliation{Novosibirsk State University, 2 Pirogova street, 630090 Novosibirsk, Russia}
\author{E.\,D.\,Gospodchikov}
\affiliation{Budker Institute of Nuclear Physics, 11 Lavrentieva prospect, 630090 Novosibirsk, Russia} 
\affiliation{Institute of Applied Physics, RAS, 46 Ulyanova street, 603950 Nizhny Novgorod, Russia}
\author{A.\,A.\,Lizunov}
\affiliation{Budker Institute of Nuclear Physics, 11 Lavrentieva prospect, 630090 Novosibirsk, Russia} 
\author{V.\,V.\,Maximov}
\affiliation{Budker Institute of Nuclear Physics, 11 Lavrentieva prospect, 630090 Novosibirsk, Russia} 
\affiliation{Novosibirsk State University, 2 Pirogova street, 630090 Novosibirsk, Russia}
\author{V.\,V.\,Prikhodko}
\affiliation{Budker Institute of Nuclear Physics, 11 Lavrentieva prospect, 630090 Novosibirsk, Russia} 
\affiliation{Novosibirsk State University, 2 Pirogova street, 630090 Novosibirsk, Russia}
\author{A.\,G.\,Shalashov}
\email{ags@appl.sci-nnov.ru}
\affiliation{Budker Institute of Nuclear Physics, 11 Lavrentieva prospect, 630090 Novosibirsk, Russia} 
\affiliation{Institute of Applied Physics, RAS, 46 Ulyanova street, 603950 Nizhny Novgorod, Russia}
\author{E.\,I.\,Soldatkina} 
\author{A.\,L.\,Solomakhin}
\affiliation{Budker Institute of Nuclear Physics, 11 Lavrentieva prospect, 630090 Novosibirsk, Russia} 
\affiliation{Novosibirsk State University, 2 Pirogova street, 630090  Novosibirsk, Russia}
\author{D.\,V.\,Yakovlev}
\affiliation{Novosibirsk State University, 2 Pirogova street, 630090 Novosibirsk, Russia}

\date{November 18, 2014}

\begin{abstract}
We demonstrate  plasma discharges with extremely high temperature of bulk electrons at the large axially symmetric magnetic mirror device GDT (Budker Institute, Novosibirsk). According to Thomson scattering measurements, the on-axis electron temperature averaged over several sequential shots is $660\pm 50$\;eV with peak values exceeding 900\;eV in few shots. This corresponds to at least threefold increase as compared to previous experiments both at the GDT and at other comparable machines, thus demonstrating the maximum quasi-stationary ($\sim1$\;ms)  electron temperature achieved in open traps. The breakthrough is made possible with application of sophisticated electron cyclotron resonance heating in addition to standard heating by neutral beams. The reported increase of the electron temperature along with previous experiments, which demonstrated high-density plasma confinement with $\beta\approx 60$\%, provide a firm basis for extrapolating to fusion relevant applications of open magnetic systems.
\end{abstract}


\maketitle

\section{Introduction}

Open magnetic systems for plasma confinement have a number of potential advantages when considered as a basis for fusion reactors with various thermonuclear applications starting from neutron sources with a thermonuclear gain factor $Q < 1$ \cite{b1,b2} and ending with power plants with $Q\gg 1$ \cite{b3,b4}. In addition to simplicity of their design, the advantages of open systems are proven capability of high-$\beta$ plasma confinement ($\beta$ is a ratio between the plasma and magnetic field pressures), relatively low wall loading by plasma heat and radiation, and possibility of direct conversion of plasma ``exhaust'' to electricity. 

The hot ion component with energy optimal for fusion  applications is commonly sustained  in such systems by injection of high-power neutral beams (NBI). In turn, the electron plasma component is heated due to relaxation of NBI driven energetic ions. Electrons with their superior mobility are carrying most of the heat flux, which goes mainly along the magnetic field lines and hits the end plates outside the magnetic mirrors. Due to direct contact with cold wall the electron temperature ($T_e$) ends up essentially lower than the mean energy of fast ions. An energy confinement time of fast ions in plasma with relatively cold electrons is determined by the electron-ion Coulomb collisions, $\tau_\mathrm{h}\propto T_e^{3/2}$. For this reason, the electron temperature is the main factor limiting the confinement time of fast ions and thus  the power efficiency of a beam-driven fusion reactor based on a magnetic mirror. 

Widely believed estimation based on classical (Spitzer) electron thermal conductivity shows that the heat flux along the magnetic field lines is proportional to $T_e^{7/2}$ \cite{b5}, which closes the door on any possible thermonuclear application of a mirror trap due to a poor quality of energy confinement. This simplified view is widely spread over the fusion community. That in combination to many experiments which never demonstrated the electron temperature higher than 280 eV \cite{b6}, led to a judgment that the very concept of a thermonuclear reactor based on a magnetic mirror has hit a dead end. This results in a relatively weak research activity in this field.

However, there are many complex and not yet sufficiently investigated processes in a region with expanding magnetic field lines behind the magnetic mirrors which lead to a significant suppression of the longitudinal electron heat flux \cite{b7}. Low-temperature experiments also suggest the inconsistency of the above mentioned simplified view \cite{b8}. This alone lets us to consider the magnetic mirror concept as a productive approach to a fusion reactor. Having said that, a direct experimental demonstration of the electron temperature much higher than that predicted by the Spitzer heat conductivity might become the strongest argument in favor of mirror trap perspectives. 

A more practical goal is to achieve a temperature required for a magnetic-mirror-based source of fusion neutrons for material testing, subcritical fission reactors and nuclear waste processing based on fusion driven burning of minor actinides. According to \cite{b2}, the electron temperature of $\approx700$\;eV is enough to fit all mentioned applications.

In this letter we report the latest experimental results, which in our opinion let us practically achieve these ambitious goals. Experiments were performed on the GDT (gas dynamic trap) facility in Budker Institute. Plasma was supported by 5\;MW NBI heating and 0.7\;MW electron cyclotron resonance heating (ECRH). In a series of successive shots with combined heating we have reached an average electron temperature of $660\pm 50$\;eV with plasma density being $0.7\times 10^{19}$\;m$^{-3}$; moreover, in several shots we have registered the electron temperature close to 1 keV. 

\section{The Gas Dynamic Trap}
The GDT is large-scale axially symmetric magnetic mirror device with 7-m-long central cell and two expander cells attached to it from both ends. Ratio between maximal and minimal confining magnetic field on the trap axis (the mirror ratio) is 35. The device is distinguished with a fairly good confinement that allows to reach local plasma $\beta$ up to $60$\%. Design, physics of plasma confinement and main goals of this machine are described in \cite{b9}; here we mention only key issues that made the reported results possible. 

Important feature of the setup is that the plasma facing collectors are placed sufficiently far from the magnetic mirrors in a region with expanded magnetic field lines. Plasma expansion before end plates essentially reduces the electron heat transfer along the magnetic field far below the Spitzer value \cite{b7,b8}. 

The main heating system consists of eight NBIs providing 5 MW of total injected power. Recently the GDT has been upgraded with the additional electron cyclotron resonance heating (ECRH) system operated at 54.5 GHz with total power 0.7 MW \cite{b10}. An evident attractive feature of the ECRH is that power is directly deposited into the electron component resulting in comparable powerflux to those transmitted to electrons from the slowing-down of NBI-born fast ions ($\sim 1$\;MW). Note, that although ECRH techniques are well developed for toroidal fusion devices and small open traps, none of the existing schemes is suitable for the GDT conditions. Thus we have developed a new scheme of the ECRH at the fundamental harmonic of the extraordinary plasma mode. In this scheme the microwave beam is launched obliquely into the plasma at a specific angle and position such that it is effectively trapped by an inhomogeneous plasma ``waveguide'', and after several crossings of the plasma axis the beam finally reaches the cyclotron resonance surface and dissipates \cite{b11}.

Presently available power supply provides the required for the ECRH magnetic field only at one end of the machine, which would limit the experiment to one of two available heating beams (with power about 0.4\;MW). Experiments in such conditions resulted in the on-axis electron temperature of $400-450$\;eV \cite{b12}. In this letter we report the experiment in which we adopt another strategy.  The confining magnetic field is reshaped such that it is increased in the both ECRH regions located near the trap ends and is lowered everywhere else. This allows exploiting both microwave beams and boosting the injected ECRH power to 0.7\;MW. Another important feature of new magnetic geometry is the possibility to tune  the magnetic field inside the ECRH region. However the whole magnetic field in the device is decreased (e.g. from 0.35 to 0.27\;T in the central solenoid and proportionally in magnetic mirrors) what results in some degradation of plasma confinement properties. In particular, without ECRH the on-axis electron temperature is only $120-180$\;eV depending on plasma density while for the standard magnetic configuration this temperature is 250 eV with plasma density $2\times 10^{19}$\;m$^{-3}$ \cite{b13}.

Another issue to be addressed is the magneto-hydro-dinamic (MHD) stability of bulk plasma, which in case of the GDT is inherently unstable. To stabilize discharge we employ the novel vortex confinement technique \cite{b14}. In short, we apply a biasing potential to the plasma edge to induce a radial electric field inside the plasma column; then $\mathbf{E}\!\times\!\mathbf{B}$\:-drift results in differential rotation of peripheral plasma which suppresses the radial transport caused by the flute MHD modes. Potential gap between plasma core and periphery should be about $T_e$ in order to provide an optimal confinement. This condition is easily violated when ECRH results in fast rise of electron temperature. Thus to stabilize plasma we boost the bias potential by applying additional voltage to the vortex confinement electrodes during the ECRH phase. 

Electron temperature and density profiles are measured in the central plane by  Thomson scattering diagnostics based on 1\;$\mu$m laser.

\section{Experimental results}
\begin {figure}[tb]
\includegraphics [width=85mm] {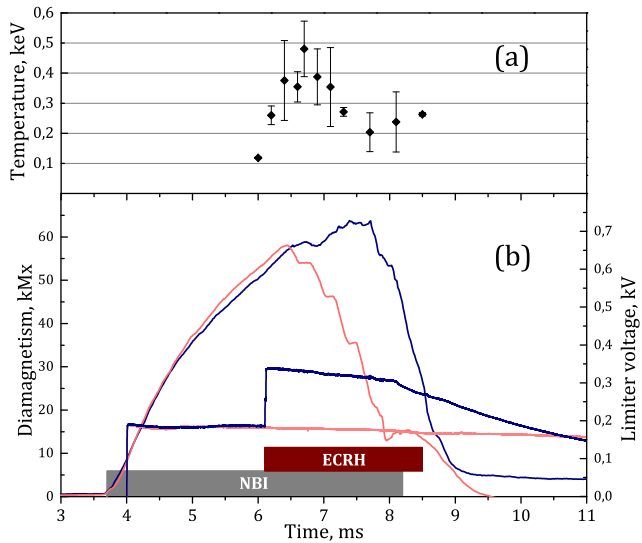}
\caption { \label{fig1} (a) Evolution of the on-axis electron temperature during the ECR heating measured using Thomson scattering system. (b) Evolution of the diamagnetic signal (mostly contributed by the hot ions) and the bias potential used for plasma stabilization during combined 5\;MW NBI and 0.7\;MW ECRH discharges with (navy lines online) and without (red lines online) additional voltage on the limiter at the GDT facility. }
\end {figure}

The experiment is performed in deuterium plasma. A typical shot is presented in Fig.\,\ref{fig1}. The discharge is initiated by a plasma gun that injects primary plasma along the trap axis from 0.5 to 4.5 ms. Plasma heating  starts at 3.7 ms with 5\;MW NBI, then after 2.4 ms an additional 0.7\;MW ECRH is switched-on. Total discharge duration is about 9 ms while the ECRH phase lasts 2.5 ms. One can see that ECRH may result in rapid degradation of plasma confinement---steady heating is switched to a full-scale instability leading to the loss of the entire plasma which is indicated by all diagnostics (most clearly by the diamagnetic loops). Nevertheless, the discharge may be stabilized by appling an additional  biasing voltage ($\approx150$\;V) to the plasma limiter during the ECRH phase. As seen from diamagnetic signals, the additional voltage stabilizes plasma, but the instability is not entirely suppressed. The latter is most clearly pronounced in time resolved measurements of the on-axis electron temperature shown in Fig.\,\ref{fig1}(a). Though the diamagnetic signal is maintained at the same level or even grows, the peaked temperature profile cannot be supported during the whole ECRH pulse. Nevertheless, the duration of stable microwave heating has been increased up to 0.6 ms what results in the record electron temperature registered at the GDT to date. Even though the MHD activity is still present, it no longer leads to the dramatic loss of the entire plasma and tends to become less destructive with the increase of the limiter potential jump.

\begin {figure}[tb]
\includegraphics [width=85mm] {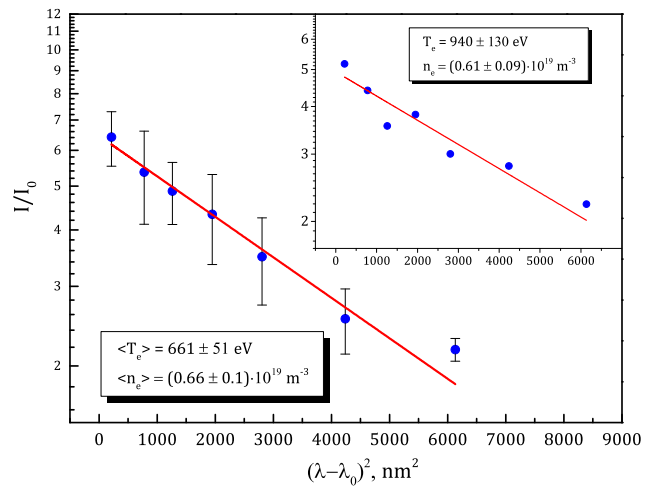}
\caption { \label{fig2} Electron energy spectrum measured by Thomson scattering on the axis of GDT and averaged over 7 consecutive shots. Fit of these data suggests the Maxwellian electron distribution function with the electron temperature of $660\pm 50$\;eV and density $(0.66\pm 0.10)\times 10^{19} $\;m$^{-3}$. The same data for one of   shots with electron temperature above 900\;eV is shown in the insert.  }
\end {figure}

In Fig.\,\ref{fig2} we present Thomson scattering data for the on-axis electron temperature. This plot is obtained by averaging over a series of several sequential shots. The scattered spectrum proves that the electron velocity distribution remained Maxwellian (represented by a straight line in the plot) with the average electron temperature of $660 \pm 50$\;eV. It should be noted that in few shots the measured electron temperature exceeds 900\;eV. Plasma density is about $0.7\times 10^{19}$\;m$^{-3}$ for all shots. Presented series of shots has been obtained in most favorable conditions for the electron heating found experimentally. In total there are more than 200 successful shots with the on-axis electron  above 300\;eV that have been registered during the reported campaign; their distribution over the measured  temperature is presented in Table \ref{tab}.

\begin{table}
\caption{\label{tab} Distribution of high-temperature shots in experiments with reduced magnetic field and 700\;kW ECRH.}
\begin{ruledtabular}
\begin{tabular}{cc}
On-axis electron temperature & Total number of shots \\
300 eV -- 500 eV & 165 \\
500 eV -- 700 eV & 43 \\
700 eV -- 900 eV & 8 \\
900 eV -- 1100 eV & 3 \\
\end{tabular}
\end{ruledtabular}
\end{table} 

In Fig.\,\ref{fig3} we show the typical radial profiles of the electron temperature and plasma density measured by the Thomson scattering in the central solenoid (before the ECRH and in 0.6 ms after the ECRH start up). The peaked temperature profile suggests that the microwave power is deposited inside the central region with characteristic radius $\sim 5$\;cm  leaving the peripheral plasma intact. This is in contradiction both to the initial theorecical work \cite{b11} and the low-power experiment \cite{b12} in which the microwave power is spread through the whole plasma cross-section  rather than focused in a narrow region around the machine axis. However, this may be explained by combination of two factors. First, for the new plasma configuration the ray-tracing indeed predicts that ECRH power deposition has a gap in the core plasma: 30\% of total injected power is deposited in a narrow and well localised region close to the plasma axis with $r<5$\;cm, 40\% of power goes to the peripheral plasma  with $r>20$\;cm, and the rest power is lost (not absorbed). Another reason for temperature peaking is a reduced central magnetic field in new  configuration that results in  poor confinement at the plasma periphery as compared to the standard configuration. 

\section{Discussion}
\begin {figure}[tb]
\includegraphics [width=85mm] {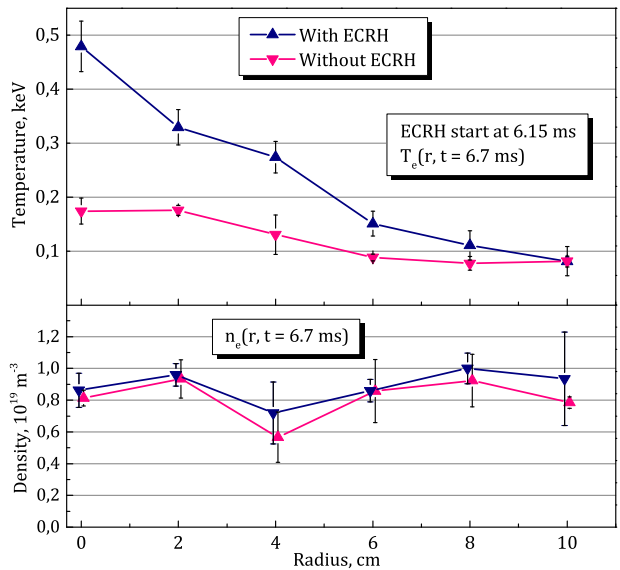}
\caption { \label{fig3} Radial profiles of the electron temperature and density in the central section of the GDT facility. Our diagnostics resolves one radial point per shot, thus these profiles were obtained as sum over series of identical shots. To improve reproducibility we use a scenario with a peak electron temperature slightly lower than that indicated in Fig.\,\ref{fig2}.}
\end {figure}

It should be stressed, however, that in spite of reduced confining properties of the machine, the power balance indicates that the core plasma is traped in the gas-dynamic regime \cite{b15}. Indeed, assuming that electron energy is lost due to plasma streaming with the ion-acoustic velocity $v_\mathrm{s}\propto T_e^{1/2}$  along the magnetic field lines, one can estimate the power density required to support a stationary discharge with a given electron temperature as  $p\propto v_\mathrm{s} T_e\propto T_e^{3/2}$. Previously this scaling has been proven experimentally for discharges without ECRH \cite{b9}. Then we can compare two discharges, before and during the stable ECRH stage, assuming that only temperature and power  are varying, 
\begin{equation*}
{T_e^\mathrm{ECRH}}/{T_e^\mathrm{NBI}}=\left(p^\mathrm{\:ECRH}/{p^\mathrm{\:NBI}}\right)^{2/3},
\end{equation*}
where $p^\mathrm{\:NBI}\!\approx\! 40$\;kW is NBI power deposited into electrons without ECRH, and  $p^\mathrm{\:ECRH}\!\approx\! 200$\;kW is the total ECRH and NBI power  after the ECRH is switched on. Both powers are calculated by ray-tracing and fast ion slowing down codes for the core plasma region limited by $r<7$\;cm (for plasma profiles shown in Fig.\;\ref{fig3}). This simple estimate gives about threefold increase in the electron temperature what is in good agreement with our  measurements. 
Note that the classical heat flux for electrons with a temperature of 500\;eV corresponds to power losses of few GWs, therefore the Spitzer thermal conductivity along magnetic field lines is suppressed in our experiment.

Summarizing our experience in ECRH supported discharge at the GDT facility we may conclude that reaching of high electron temperature in an open trap with dense plasma results in degradation of plasma MHD stability. This problem is manageable but eventually we will have to find some compromise between high temperature and stability in practical applications. In the present report we aim to demonstrate the highest possible (in a quasi-stationary discharge) electron temperature with available resources. To make it possible we focus the microwave power deposition on the plasma axis reaching very high local power density (up to 20\;kW/cm$^2$ compared to $0.1-0.3$\;kW/cm$^2$ typical of purely NBI heating).  Moreover, ray-tracing calculations reveal a positive feedback: temperature increase results in the better absorbtion and, therefore, the stronger peaking of a temperature profile.  
 
\section{Conclusion}
We find that our theoretical understanding of resonant plasma heating has been proven experimentally, thus the proposed novel ECRH scheme works quite robust. So we are quite confident on straightforward continuation of the experiment by expanding our technical limitations. Returning to the standard magnetic configuration with 0.35 T in the trap center solenoid and further increase of vortex potential will improve overall plasma stability and confinement, what hopefully will allow to operate the ECRH as a routine tool at the GDT device. The practical goal is to obtain more broad temperature profiles with high electron energy content.

The reported increase of the electron temperature along with previous experiments, which demonstrated plasma confinement with beta as high as 60\%, provide a firm basis for extrapolating the gas dynamic trap concept to a fusion relevant applications.

\begin{acknowledgments}
The authors are grateful to Prof.\,M.\,Thumm for his valuable advises concerning physical and technical aspects of ECR heating and to Dr.\,A.\,Anikeev, Dr.\,O.\,Smolyakova,  Dr.\,V.\,Malygin and V.\,Orlov for their support and assistance. We thank all the GDT team participated in the experiment.  We also appreciate encouragement and interest from Profs.\,A.\,Ivanov, A.\,Arzhannikov, D.\,Ryutov, A.\,Molvik and T.\,S.\,Simonen.
The work is supported by the Russian Scientific Foundation (pr. 14-12-01007).
\end{acknowledgments}

\end {document}